\documentclass[twoside,slac_one]{revtex4}
\usepackage{graphicx}
\usepackage{fancyhdr}
\usepackage{amsmath} 
\usepackage{bm}
\usepackage{amsxtra}
\usepackage{amssymb}
\usepackage{amsthm}
\usepackage{latexsym}
\usepackage{lscape}
\usepackage{relsize}
\RequirePackage{xspace}
\def\babar{\mbox{\slshape B\kern-0.1em{\smaller A}\kern-0.1em B\kern-0.1em{\smaller A\kern-0.2em R}}}

\newcommand{\jprlBase}       {Phys.\ Rev.\ Lett.\xspace}

\newcommand{\jprl}      [1]  {\jprlBase\ {\bf #1}}

\def\Ds       {\ensuremath{D^-_s}\xspace}
\def\Dsm       {\ensuremath{D^-_s}\xspace}

\def\Dz       {\ensuremath{D^0}\xspace}
\def\Dbar     {\kern 0.2em\overline{\kern -0.2em D}{}\xspace}
\def\Dzb      {\ensuremath{\bar{D}^0}\xspace}
\def\Dp       {\ensuremath{D^+}\xspace}
\def\Dm       {\ensuremath{D^-}\xspace}

\def\epem     {\ensuremath{e^+e^-}\xspace}

\def\ccbar    {\ensuremath{c\bar c}\xspace}

\def\pip      {\ensuremath{\pi^+}\xspace}
\def\pim      {\ensuremath{\pi^-}\xspace}

\def\KS       {\ensuremath{K^0_{\scriptscriptstyle S}}\xspace}
\def\Kp       {\ensuremath{K^+}\xspace}
\def\Km       {\ensuremath{K^-}\xspace}

\def\Kbar     {\kern 0.2em\overline{\kern -0.2em K}{}\xspace}

\def\mevcc  {MeV\ensuremath{/c^2}\xspace}

\def\pcm {\ensuremath{p^*}\xspace}

\def\Ds       {\ensuremath{D^+_s}\xspace}
\def\Dsm       {\ensuremath{D^-_s}\xspace}
\def\Dz       {\ensuremath{D^0}\xspace}
\def\Dbar     {\kern 0.2em\overline{\kern -0.2em D}{}\xspace}
\def\Dzb      {\ensuremath{\Dbar^0}\xspace}
\def\Dp       {\ensuremath{D^+}\xspace}
\def\Dm       {\ensuremath{D^-}\xspace}

\def\epem     {\ensuremath{e^+e^-}\xspace}
\def\ccbar    {\ensuremath{c\overline c}\xspace}

\def\pip      {\ensuremath{\pi^+}\xspace}
\def\pim      {\ensuremath{\pi^-}\xspace}

\def\Kp       {\ensuremath{K^+}\xspace}
\def\Km       {\ensuremath{K^-}\xspace}

\def\DstarPlusPi0 {\ensuremath{ {D}^{*+}\pi^0}\xspace}

\def\Dz      {\ensuremath{D^0}\xspace}
\def\KS    {\ensuremath{K^0_{\scriptscriptstyle S}}\xspace}

\def\Kz    {\ensuremath{K^0}\xspace}

\def\CP                {\ensuremath{C\!P}\xspace}
\def\CPV                {\ensuremath{C\!P\!V}\xspace}
\def\T                {\ensuremath{T}\xspace}
\def\Kbar  {\kern 0.2em\overline{\kern -0.2em K}{}\xspace}

\def\Kz    {\ensuremath{K^0}\xspace}
\def\Kzb   {\ensuremath{\Kbar^0}\xspace}
\def\KzKzb {\ensuremath{\Kz \kern -0.16em \Kzb}\xspace}
\def\KS    {\ensuremath{K^0_{\scriptscriptstyle S}}\xspace} 

\newcommand{\stat}{\ensuremath{\mathrm{(stat)}}\xspace}
\newcommand{\syst}{\ensuremath{\mathrm{(syst)}}\xspace}
\newcommand{\progtp}    [1]  {{Prog.\ Theor.\ Phys.\ {\bf #1}}}
\def\todd     {\ensuremath{T}-odd\xspace}
\def\DpDecay  {\ensuremath{\Dp \to \Kp \KS \pip \pim}\xspace}
\def\DsDecay  {\ensuremath{\Ds \to \Kp \KS \pip \pim}\xspace}

\def\Dps      {\ensuremath{D^+_{(s)}}\xspace}

\def\Dsm      {\ensuremath{D^-_{s}}\xspace}
\def\Ct       {\ensuremath{C_T}\xspace}
\def\Ctb      {\ensuremath{\bar{C}_T}\xspace}
\def\At       {\ensuremath{A_T}\xspace}
\def\Atbar    {\ensuremath{\bar{A}_T}\xspace}
\def\Atv      {\ensuremath{\mathcal{A}_T}\xspace}
\def\CPT      {\ensuremath{C\!PT}\xspace}
\def\T        {\ensuremath{T}\xspace}
\begin{document}

\title{Recent \boldmath{\CP} Violation Studies from \babar}

\author{R.M. White}
\affiliation{Department of Physics and Astronomy, University of South Carolina, Columbia, SC, USA}

\begin{abstract}
In this proceeding, results of searches for \CP violation in charm decays using the 
full \babar~dataset are discussed. The parameter $A_{CP}$ in the decay $D^\pm\to\KS\pi^\pm$ 
is determined to be $(-0.39 \pm 0.13 \pm 0.10)\%$. Meaurements of \CP violation
using \T-odd correlations in the four-body decays \DpDecay and \DsDecay are
$( -12.0 \pm  10.0_{\stat} \pm 4.6_{\syst} ) \times 10^{-3}$ and
$( -13.6 \pm 7.7_{\stat} \pm 3.4_{\syst} ) \times10^{-3}$, respectively.

\end{abstract}

\maketitle

\thispagestyle{fancy}

\section{Introduction}
In the Standard Model (SM), \CP violation (\CPV) arises from the complex phase of the 
CKM quark-mixing matrix~\cite{Kobayashi:1973fr}.  
Measurements of the \CPV asymmetries in 
the $K$ and $B$ meson systems are consistent with 
expectations based on the SM and, together with 
theoretical inputs, lead to the 
determination of the parameters of the CKM matrix.
\CPV has not yet been observed in the  
charm sector, where the theoretical predictions based 
on the SM for \CPV asymmetries are at the level of 
$10^{-3}$ or below~\cite{Buccella:1994nf}. An observation 
of \CP asymmetries at the level of one percent or greater would
be a clear indication of new physics.

\section{Search for \boldmath{\CP} Violation in the decay {\boldmath$D^+\to K_{S}^0\pi^+$}~\cite{DtoKspi}}
\babar~searched for \CPV in the decay $D^\pm\to\KS\pi^\pm$ by measuring the parameter $A_{\CP}$ defined as:
\begin{equation}
	A_{\CP}=\frac{\Gamma(D^+\to\KS\pi^+)-\Gamma(D^-\to\KS\pi^-)}
	{\Gamma(D^+\to\KS\pi^+)+\Gamma(D^-\to\KS\pi^-)},
\end{equation}
where $\Gamma$ is the partial decay width for this decay. 
This decay mode has been chosen because of its clean experimental signature. 
Although direct \CP violation due to interference between
Cabibbo-allowed and doubly Cabibbo-suppressed amplitudes is 
predicted to be negligible within the SM~\cite{Lipkin:1999qz}, $\Kz-\Kzb$ mixing induces a 
time-integrated \CP violating asymmetry of $(-0.332\pm 0.006)\,\%$~\cite{Nakamura:2010zzi}.
Contributions from non-SM processes may reduce the value of the measured 
$A_{\CP}$ or enhance it up to the level of one percent~\cite{Lipkin:1999qz,Bigi:1994aw}.
Therefore, a significant deviation of the $A_{\CP}$ measurement from pure $\Kz-\Kzb$
mixing effects would be evidence for the presence of new physics beyond the SM.
Due to the smallness of the expected value, this measurement requires a large data sample and precise
control of the systematic uncertainties.
Previous measurements of $A_{\CP}$ have been reported by the CLEO-c 
($(-0.6\pm 1.0 \stat \pm 0.3 \syst)\%$~\cite{:2007zt}) and
Belle collaborations ($(-0.71\pm 0.19 \stat \pm 0.20 \syst)\%$~\cite{Ko:2010ng}).

We select $D^\pm\to\KS\pi^\pm$ decays by combining a $\KS$ 
candidate reconstructed in the decay mode
$\KS\to\pi^+\pi^-$ with a charged pion candidate.
A \KS candidate is reconstructed from two oppositely charged
tracks with an invariant mass within 
$\pm$ 10 \mevcc of the nominal \KS mass~\cite{Nakamura:2010zzi}.
To obtain the final candidate events, a Boosted Decision Tree (BDT) 
algorithm~\cite{Speckmayer:2010zz} is constructed from seven 
discriminating variables for each $D^\pm$ candidate:
the measured proper decay time $\tau(D^\pm)$, 
the decay distance in the transverse plane $L_{xy}(D^\pm)$,
the CM momentum magnitude $p^*(D^\pm)$, the momentum magnitudes and transverse components 
with respect to the beam axis for both the \KS and pion candidates.

\begin{figure}[tb]
\begin{center}
\includegraphics[width=8cm]{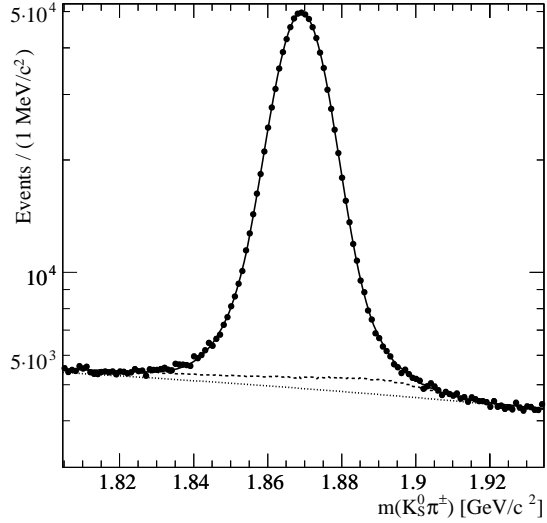}
\vspace{-0.3cm}
\caption{Invariant mass distribution for $\KS \pi^\pm$ candidates in the data (black points).
The solid curve shows the fit to the data. The dashed line is the sum of all backgrounds,
while the dotted line is combinatorial background only.	The vertical scale of the plot is logarithmic.}
\label{dtokspi}
\vspace{-0.7cm}
\end{center}
\end{figure}

\begin{figure}[tb]
\begin{center}
\includegraphics[width=12cm]{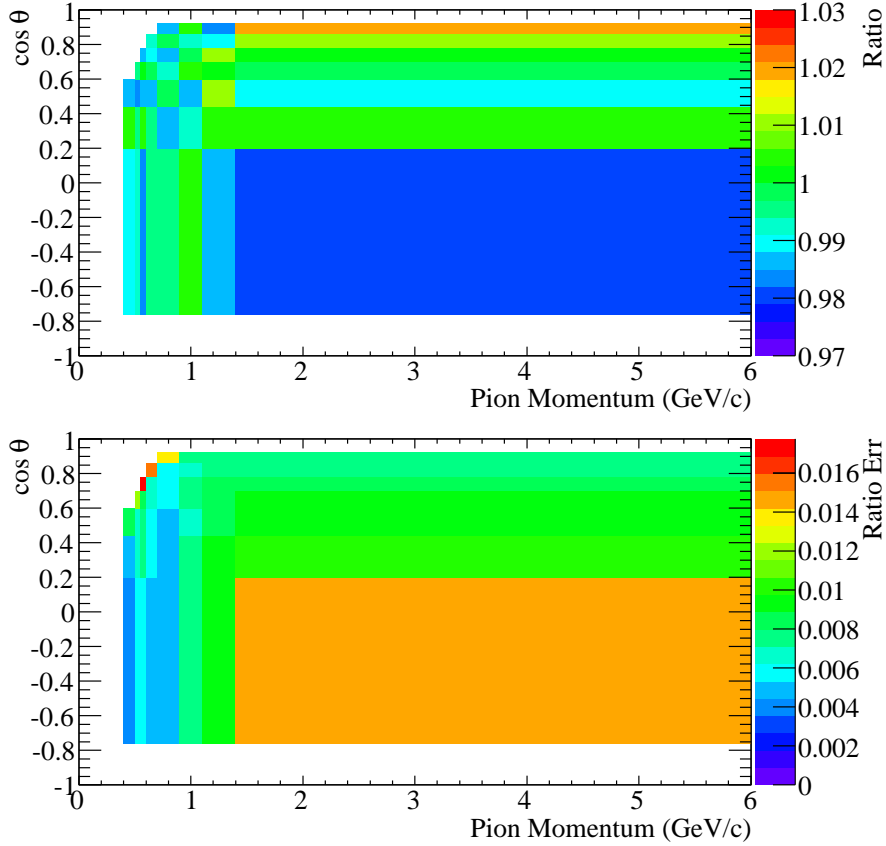}
\vspace{-0.3cm}
\caption{Map of the ratio between detection efficiency
for $\pi^+$ and $\pi^-$ (top) plus the
corresponding statistical errors (bottom). 
    The map is produced using the numbers of $\pi^-$ and $\pi^+$ 
    tracks in the selected control sample.}
    \label{trackratio}
    \vspace{-0.7cm}
    \end{center}
\end{figure}

A binned maximum likelihood (ML) fit to the $m(\KS \pi^\pm)$ distribution for the
retained $D^\pm$ candidates is used to extract the signal yield. 
The total probability distribution function (PDF) is the sum of signal and background components. The signal
PDF is modeled as a sum of three Gaussian functions, the first two of them with common mean.
The background PDF is taken as a sum of
two components: a background from 
$D^\pm_s\to\KS K^\pm$, where the $K^\pm$ is
misidentified as $\pi^\pm$, and a combinatorial background from other sources.
The data and the fit are shown in Fig.~\ref{dtokspi}.
All of the fit parameters are extracted from the fit to the data sample 
apart from the normalization of the background due to $D^\pm_s\to \KS K^\pm$, 
      which is fixed to the value predicted by the MC simulation.
We determine $A_{\CP}$ by measuring the signal yield asymmetry $A$ defined as:
\begin{equation}
A=\frac{N_{D^+}-N_{D^-}}{N_{D^+}+N_{D^-}},
\end{equation}
where $N_{D^+}$($N_{D^-}$) is the number of fitted 
$D^+\to\KS\pi^+$($D^-\to\KS\pi^-$) decays.
The quantity $A$ is the result of two other contributions in addition to $A_{\CP}$. 
There is a physics component due to the forward-backward (FB) asymmetry ($A_{FB}$)
in $\epem\to\ccbar$, arising from $\gamma^*$-$Z^0$ interference and 
high order QED processes in $\epem \to \ccbar$. This asymmetry
will create a difference in the number of reconstructed $D^+$ and $D^-$ 
decays due to the FB detection asymmetries arising from the boost of the 
center-of-mass (CM) system relative to the laboratory frame.  
There is also a detector-induced component 
due to the difference in the reconstruction efficiencies of $D^+\to 
K^0_s\pi^+$ and $D^-\to K^0_s\pi^-$ generated by 
differences in the track reconstruction and identification efficiencies 
for $\pi^+$ and $\pi^-$. While $A_{FB}$ is measured together with $A_{\CP}$
using the selected dataset, we correct the dataset itself for the
reconstruction and identification effects using control data sets.
\babar~developed a data-driven method to determine the 
charge asymmetry in track reconstruction as a function of the magnitude of 
the track momentum and its polar angle which is shown along with the
associated errors in Fig.~\ref{trackratio}.

Neglecting the second-order terms that contain the product of $A_{\CP}$ and $A_{FB}$,
the resulting asymmetry can be expressed simply as the sum of the two.
The parameter $A_{\CP}$ is independent of kinematic variables,
while $A_{FB}$ is an odd function of $\cos\theta^*_D$, 
where $\theta^*_D$ is the polar angle of the $D^\pm$ 
candidate momentum in the $\epem$ CM frame.
If we compute $A(+|\cos\theta^*_D|)$ for the $D^\pm$ candidates 
in a positive $\cos\theta^*_D$ bin and $A(-|\cos\theta^*_D|)$ for 
the candidates in its negative counterpart,
the contribution to the two asymmetries from $A_{\CP}$ is the same, 
while the contribution from $A_{FB}$ has the same magnitude but opposite sign.
Therefore $A_{\CP}$ and $A_{FB}$ can be written as a function of $|\cos\theta^*_D|$ as follows:
\begin{align}
A_{FB}(|\cos\theta^*_D|) &= \frac{A(+|\cos\theta^*_D|) - A(-|\cos\theta^*_D|)}{2} \\ 
\intertext{and}
A_{\CP}(|\cos\theta^*_D|) &= \frac{A(+|\cos\theta^*_D|) + A(-|\cos\theta^*_D|)}{2}.
\label{eq:AcpAfb_intro}
\end{align}

\begin{figure}[tb]
	\begin{center}
	\includegraphics[width=0.4\textwidth,clip=true]{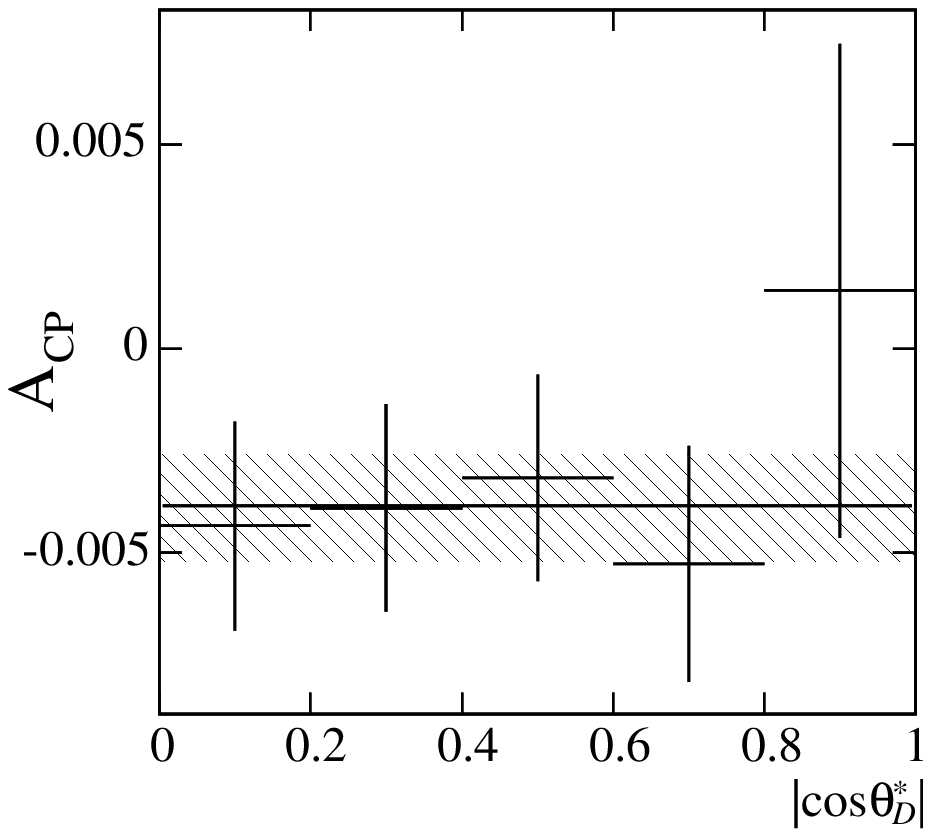}
	\includegraphics[width=0.4\textwidth,clip=true]{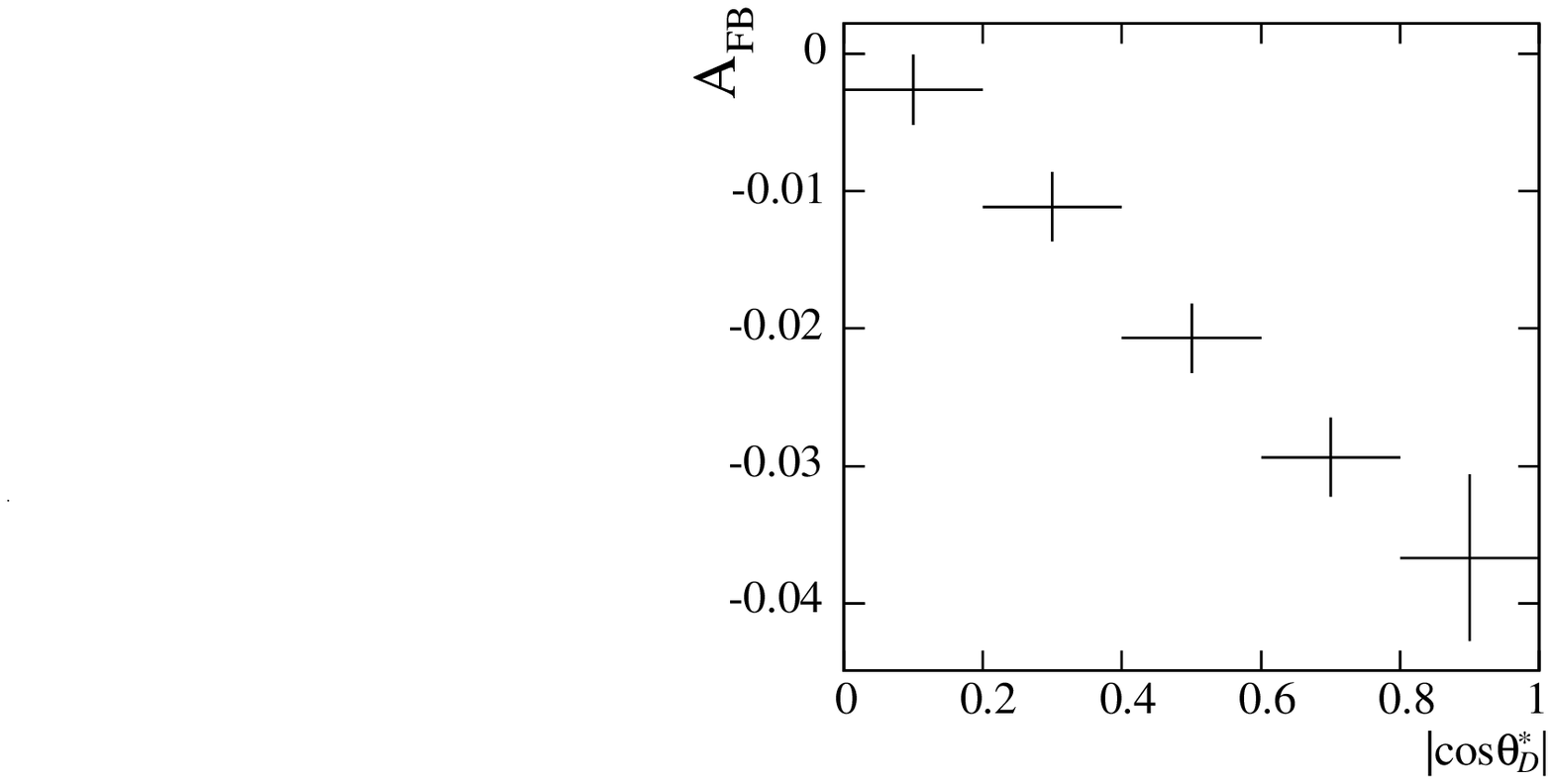}
	\vspace{-0.3cm}
	\caption{$A_{\CP}$ (top) and $A_{FB}$ (bottom) asymmetries for $D^\pm\to\KS\pi^\pm$ candidates
	    as a function of $|\cos\theta^*_D|$ in the data sample. The solid line represents the central value of
	    $A_{\CP}$ and the hatched region is the $\pm1\,\sigma$ interval, both obtained from a $\chi^2$ minimization
	    assuming no dependence on $|\cos\theta^*_D|$.}
	\label{acp}
	\vspace{-0.7cm}
	\end{center}
\end{figure}

The selected sample is divided into ten subsamples corresponding to
ten $\cos\theta^*_D$ bins of equal width and a simultaneous binned 
ML fit is performed on the invariant mass distributions of $D^+$ and 
$D^-$ candidates for each subsample to extract the signal yield asymmetries.
Using the asymmetry measurements in five positive and in five negative 
$\cos\theta^*_D$ bins, we obtain five $A_{FB}$ and five $A_{CP}$ values. 
As $A_{CP}$ does not depend upon $\cos\theta^*_D$, we compute 
a central value of this parameter using a $\chi^2$ minimization
to a constant. The $A_{\CP}$ and $A_{FB}$ values are shown in Fig.~\ref{acp}, together with
the central value and $\pm 1\,\sigma$ confidence interval for $A_{\CP}$. We determine
$A_{CP}$ to be:
\begin{equation}
A_{\CP}=(-0.39 \pm 0.13 \pm 0.10)\%
\end{equation}
where the first error is statistical and the second systematic.
	
\section{Search for {\boldmath\CP} Violation using {\boldmath\T}-Odd Correlations in {\boldmath$D^+_{(s)}\rightarrow K^0_S K^+\pi^+\pi^+$}~\cite{delAmoSanchez:2011fb}}
A search for \CP violation in the decays \DpDecay and \DsDecay using \todd correlations is described here.
We define a kinematic triple product that is odd under time reversal using the vector momenta of the final state particles in the \Dps rest frame as
\begin{equation}
\Ct \equiv \vec{p}_{\Kp} \cdot \left( \vec{p}_{\pip} \times \vec{p}_{\pim} \right).
\label{eq:Ct}
\end{equation}
Under the assumption of \CPT invariance, \T violation is equivalent to \CP violation.

We study the \todd correlations by measuring the observable expressed in Eq.~(\ref{eq:Ct}) and then evaluating the asymmetry
\begin{equation}
\At \equiv \frac{\Gamma(\Ct>0) - \Gamma(\Ct<0)}{\Gamma(\Ct>0) + \Gamma(\Ct<0)},
\label{eq:At}
\end{equation}
where $\Gamma$ is the decay rate for the process under study.
The observable defined in Eq.~(\ref{eq:At}) can have a non-zero value due to final state interactions even if the weak phases are zero~\cite{Bigi:2009zzb}.
The \todd asymmetry measured in the \CP-conjugate decay process, \Atbar, is defined as: 
\begin{equation}
\Atbar \equiv \frac{\Gamma(-\Ctb>0) - \Gamma(-\Ctb<0)}{\Gamma(-\Ctb>0) + \Gamma(-\Ctb<0)},
\label{eq:Atb}
\end{equation}
where $\Ctb\equiv \vec{p}_{\Km} \cdot \left( \vec{p}_{\pim} \times \vec{p}_{\pip} \right)$. 
We can then construct:
    \begin{equation}
    \Atv \equiv \frac{1}{2}\left( \At - \Atbar \right),
    \label{eq:Atv}
    \end{equation}
which is an asymmetry that characterizes \T violation in the weak decay process~\cite{Bensalem:2002ys,Bensalem:2002pz,Bensalem:2000hq}.

At least four different particles are required in the final state so that the triple product may be defined using momentum vectors only~\cite{Golowich:1988ig}.
The $D$ meson decays suitable for this analysis method are \DpDecay, \DsDecay and $\Dz\to\Kp\Km\pip\pim$. 
The search for \CP violation using \todd correlations in $\Dz\to\Kp\Km\pip\pim$ has recently been carried out by the \babar\  Collaboration, 
and no evidence of \CP violation has been observed~\cite{delAmoSanchez:2010xj}.

\begin{figure}
\centering
\includegraphics[width=6cm]{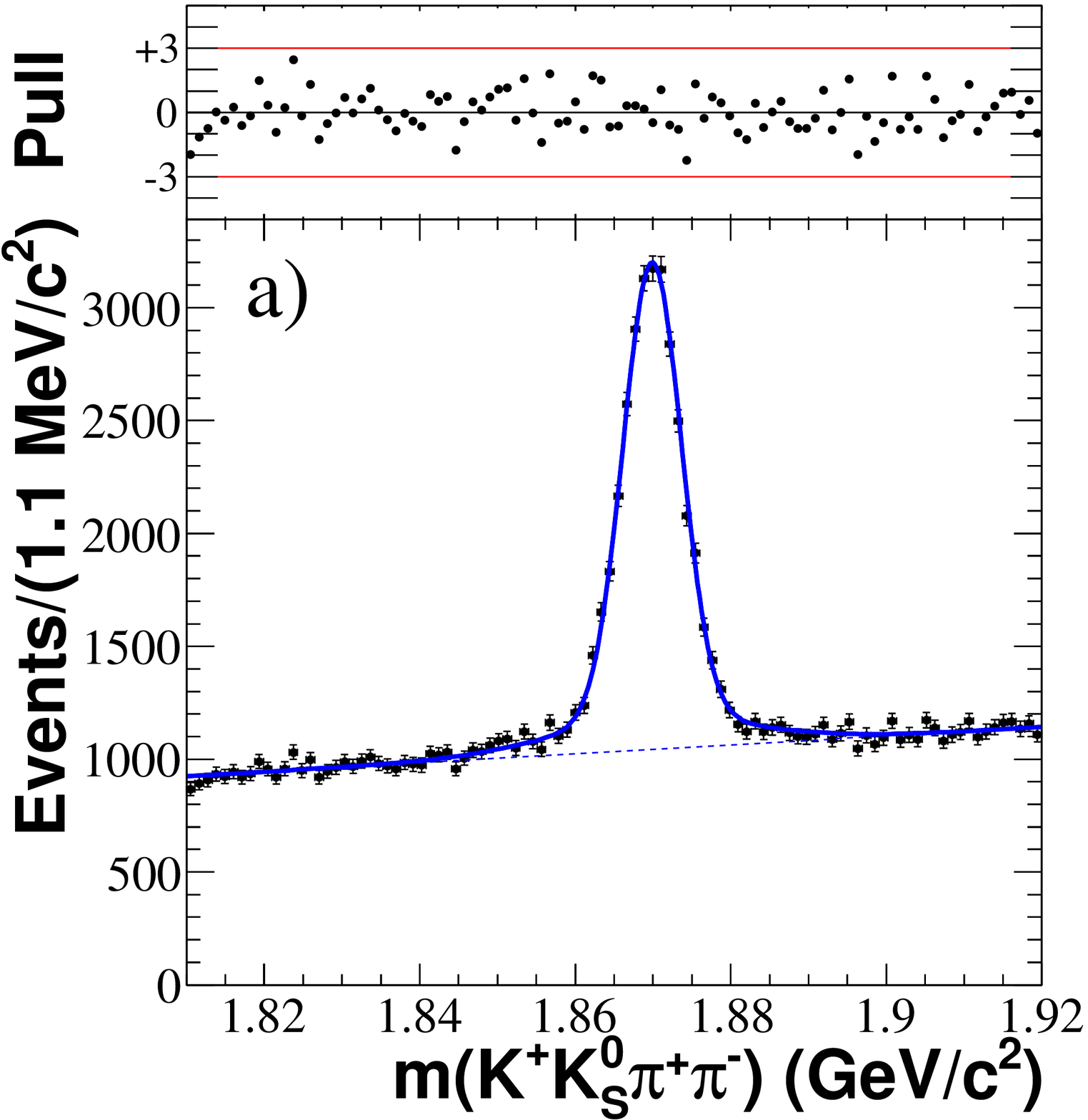}
\includegraphics[width=6cm]{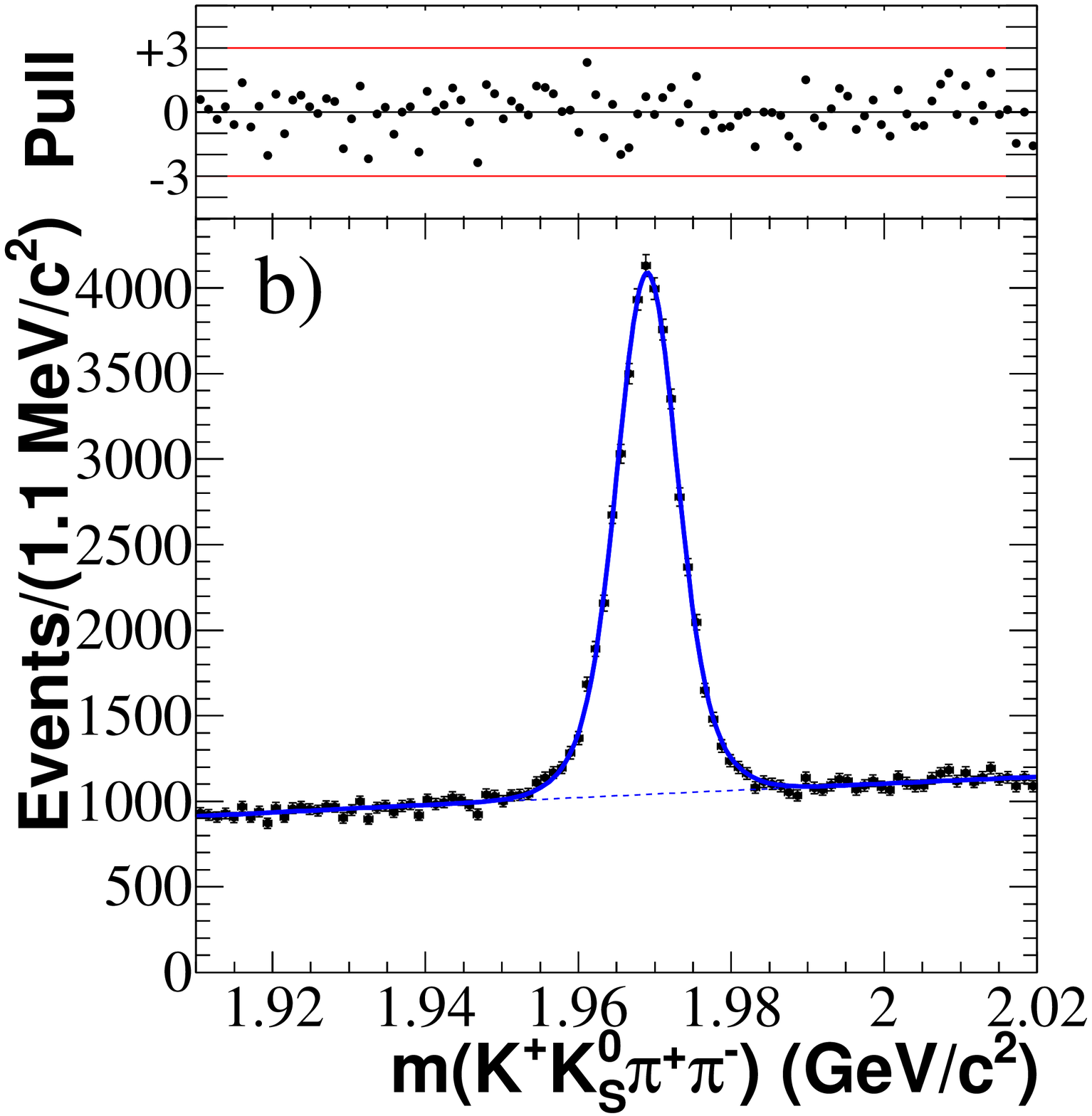}
\caption{\label{fig:fig2} 
The $\Kp\KS\pip\pim$ mass spectrum a) in the \Dp, and b) in the \Ds mass region.
    The curves result from the fits described in the text. The distributions of the Pull values are also shown.}
\end{figure}

The \Dp and \Ds meson decay candidates are reconstructed in the production and decay sequence:
\begin{equation}
\epem\to X\Dps; \Dps\to\Kp\KS\pip\pim; \KS\to\pip\pim,
\label{eq:reaction}
\end{equation}
using the events with at least five charged particles. To obtain the final set of signal candidates, the \pcm, the difference
in vertex probabilities that the parent meson originates from a common vertex and the primary vertex, and the signed transverse
decay length are combined in a likelihood-ratio test.
Fig.~\ref{fig:fig2} shows the resulting $\Kp\KS\pip\pim$ mass spectra in the \Dp and \Ds regions.
For each region, the signal is described by the superposition of two Gaussian functions with a common mean value.
The background is parametrized by a first-order polynomial in the \Dp region, and by a second-order polynomial in the \Ds region. 
We extract the integrated yields $N(\Dp) = 21210 \pm 392$ and $N(\Ds) = 29791 \pm 337$ from the fits, where the uncertainties are statistical only.

We next divide the data sample into four sub-samples depending on $D_{(s)}$ charge and whether \Ct (\Ctb) is greater or less than zero,
and fit the corresponding mass spectra simultaneously to extract the yields and the values of the asymmetry parameters \At and \Atbar.
The triple product asymmetries for Cabibbo-suppressed decays $\Dz\to K^+K^-\pi^+\pi^+$~\cite{delAmoSanchez:2010xj}, 
\DpDecay and Cabibbo-favored decays \DsDecay are summarized in Tab.~\ref{tab:todd}. The average of the triple product asymmetries
is also included in the table
\begin{equation}
\Sigma_T = \frac{1}{2}(\At + \Atbar)
\end{equation}
which is not a CP violating parameter but may provide more information on the final-state interactions in these decays.

\begin{table}[h]
\begin{center}
\caption{Triple-product asymmetries \At, \Atbar, \Atv, and $\Sigma_T$ for the Cabibbo-suppressed decays 
$\Dz\to K^+K^-\pi^+\pi^-$~\cite{delAmoSanchez:2010xj}, \DpDecay~\cite{delAmoSanchez:2011fb} and the Cabibbo-favored decays 
\DsDecay~\cite{delAmoSanchez:2011fb}. The values quoted in units $10^{-3}$.}
\begin{tabular}{|l|c|c|c|}
\hline 
\hline
Asymmetry & \Dz/\Dzb & \Dp/\Dm & \Ds/\Dsm \\
\hline
\At & -68.5 $\pm$ 7.3 $\pm$ 5.8 & 11.2 $\pm$ 14.1 $\pm$ 5.7 & -99.2 $\pm$ 10.7 $\pm$ 8.3 \\
\Atbar & -70.5 $\pm$ 7.3 $\pm$ 3.9 & 35.1 $\pm$ 14.3 $\pm$ 7.2 & -72.1 $\pm$ 10.9 $\pm$ 10.7 \\
\Atv & 1.0 $\pm$ 5.1 $\pm$ 4.4 & -12.0 $\pm$ 10.0 $\pm$ 4.6 & -13.6 $\pm$ 7.7 $\pm$ 3.4 \\
$\Sigma_T$ & -69.5 $\pm$ 6.2 & 23.1 $\pm$ 11.0 & 85.6 $\pm$ 10.2 \\    
\hline
\hline
\end{tabular}
\label{tab:todd}
\end{center}
\end{table}

The final measurements for \Atv in all decays are consistent with zero, however, the values for the \T-odd asymmetries are considerably
larger in \Dz and \Ds decays. The differences in these values for the various decays may indicate a difference in the final-state
interactions. The final-state interactions may be responsible for the hierarchy of lifetimes and branching fractions~\cite{GronauRosner}.

\section{Conclusion}
Measurements with the final \babar~dataset achieve the precision at the SM prediction for \CP violation in charm decays. The 
systematic uncertainties are at the level of the statistical uncertainties. Current and future measurements from LHCb, 
Belle, and SuperB will face the challenge of reducing these systematic uncertainties. 

\bigskip 

\end{document}